\documentclass[prb,twocolumn,showpacs,preprintnumbers,amsmath,amssymb]{revtex4}

\usepackage{graphicx}
\usepackage{dcolumn}
\usepackage{bm}

\begin{document}


\title{Thermodynamic and thermoelectric properties of high-temperature cuprate superconductors in the stripe phase}

\author{T. Tohyama}
\affiliation{Institute for Materials Research, Tohoku University, Sendai 980-8577,
Japan}
\author{S. Maekawa}
\affiliation{Institute for Materials Research, Tohoku University, Sendai 980-8577,
Japan}
\author{P. Prelov\v sek}
\affiliation{Faculty of Mathematics and Physics, University of Ljubljana,
and J. Stefan Institute, 1000 Ljubljana, Slovenia}

\date{\today}

\begin{abstract}
We examine the thermodynamic and thermoelectric properties in the stripe
phase of high-$T_c$ cuprates, by using the finite-temperature Lanczos technique for
the $t$-$J$ model with a potential that stabilizes vertical charge stripes.
When the stripe potential is turned on, the entropy is suppressed as a consequence
of the formation of one-dimensional charge stripes accompanied by an enhancement
of antiferromagnetic spin correlation in the spin domains.  The stripe formation
leads also to weak temperature dependence of the chemical potential, leading to
the suppression of the thermoelectric power.  The suppression of the entropy
and thermoelectric power is consistent with experimental data in the stripe
phase of La$_{1.6-x}$Nd$_{0.4}$Sr$_x$CuO$_4$.
\end{abstract}

\pacs{74.25.Bt, 71.10.Fd, 74.25.Fy, 74.72.Dn}
\maketitle

Spin and charge ordering in high-$T_c$ cuprates has been reported in
neutron-scattering experiments of
La$_{1.475}$Nd$_{0.4}$Sr$_{0.125}$CuO$_4$ (Ref.~1) and
La$_{1.875}$(Ba,Sr)$_{0.125}$CuO$_4$ (Ref.~2) at the hole concentration
$x=1/8$.  The ordered phase is called the stripe phase, because the ordering is
explained by a stripe structure that consists of one-dimensional (1D) charge domain
walls with the filling of 1/2 hole per site (quarter filling) separating
antiferromagnetic (AF) spin domains being in antiphase.~\cite{Tranquada}
The presence of the quarter-filled
charge river has been confirmed by angle-resolved photoemission spectroscopy
experiments for La$_{1.28}$Nd$_{0.6}$Sr$_{0.12}$CuO$_4$.~\cite{Zhou}  At around
$x\sim1/8$ the superconducting transition temperature $T_c$ is strongly
suppressed, which is known as the 1/8 anomaly.

The stripe phase emerges below the temperature of the structural phase transition
from the low-temperature orthorhombic (LTO) to tetragonal (LTT)
phase.~\cite{Tranquada,Fujita}  Transport properties exhibit anomalous behaviors
in the stripe phase with LTT structure.  For example, the Hall constant $R_H$ becomes
strongly temperature dependent in the stripe phase and it eventually almost vanishes,
$R_H \sim0$ at temperature $T\sim 0$.~\cite{Nakamura,Noda,Arumugam,Adachi}  The vanishing
$R_H$ is interpreted as a particular consequence of strong correlation
in the stripe structure where the formation of the 1D charge stripe with
an equal concentration of hole and electron carriers (due to the strong on-site
Coulomb interaction) leads to the particle-hole
symmetry and thus the vanishing of the off-diagonal conductivity
$\sigma_{xy}$.~\cite{Emery,Prelovsek}

Below the LTO-LTT transition temperature, the thermoelectric power $Q$ also shows
a sudden decrease similar to $R_H$ at around $x=1/8$ and becomes negative
at $x=1/8$.~\cite{Nakamura,Adachi,Sera,Takeda}  In addition, the electronic
specific heat coefficient $\gamma$ at $x=1/8$ is very small at low temperatures.~\cite{Takeda}
These thermodynamic and thermoelectric anomalies are likely to be related
to the formation of static stripe.  In the present
paper, we investigate thermodynamic properties such as the chemical potential $\mu$ and
the entropy $s$ in the stripe phase of cuprates, and clarify the origin
of the suppression of $Q$ as well as $s$ observed in the experiments.
We employ in this study the $t$-$J$ model with
a stripe potential that stabilizes the charge stripe, and use the
finite-temperature Lanczos method~\cite{Jaklic} for small clusters to calculate the
temperature dependence of the thermodynamic properties.
We find that the entropy $s$ is suppressed in the stripe phase as a consequence of
the formation of the 1D charge stripes accompanied by the development
of AF spin domains.  Such a stripe formation simultaneously leads to weak temperature
dependence of $\mu$.  The thermoelectric power obtained from the temperature
dependence of $\mu$ becomes small and negative under the strong stripe potential.
The suppression of $s$ and $Q$ is consistent with experimental data in the
stripe phase of La$_{1.6-x}$Nd$_{0.4}$Sr$_x$CuO$_4$.

In order to study the thermodynamics of the stripe phase of cuprates, we use the
finite-temperature Lanczos technique within the grand-canonical
ensemble~\cite{Jaklic} for the $t$-$J$ model with a stripe potential.  The prototype
$t$-$J$ model reads
\begin{equation}
H_{tJ}=-t\sum_{\langle ij\rangle s}\left( \tilde{c}^\dagger_{js}\tilde{c}_{is}+
{\rm H.c.} \right)+J\sum_{\langle ij\rangle} \left( {\bf S}_i\cdot {\bf S}_j -
\frac{1}{4} n_i n_j \right), \label{HtJ}
\end{equation}
where no double occupancy of sites is allowed. In numerical studies we use $J/t=0.4$ to
be in the strong-correlation regime relevant to cuprates.  The stripes are stabilized
by introducing an attractive hole potential $V$ along the
stripe,~\cite{Prelovsek,Tohyama,Shibata,Riera}
\begin{equation}
H_\mathrm{st}=-V\sum_{i \in \mathrm{stripe}} (1-n_i),
\label{HV}
\end{equation}
which on one hand in a small system enhances the intrinsic tendency of the
$t$-$J$ model towards the stripes structures~\cite{White} and on the other
hand represents the effects of extrinsic features favorable to the formation
of the stripes such as lattice anisotropy.~\cite{Kampf}

We consider here numerically two types of $t$-$J$ clusters with $N=N_x\times N_y$ sites:
(i) $(N_x,N_y)$=(4,4) with periodic boundary conditions in both the $x$ and
$y$ directions and (ii) $(N_x,N_y)$=(4,5) with a periodic boundary condition
in the $x$ direction and an open boundary condition in the $y$ direction.
The stripe potential Eq.~(\ref{HV}) is introduced into a row along the
$x$ direction of the $4\times 4$ cluster and into a middle leg of
the $4\times 5$ cluster.

\begin{figure}
\begin{center}
\includegraphics[width=8.cm]{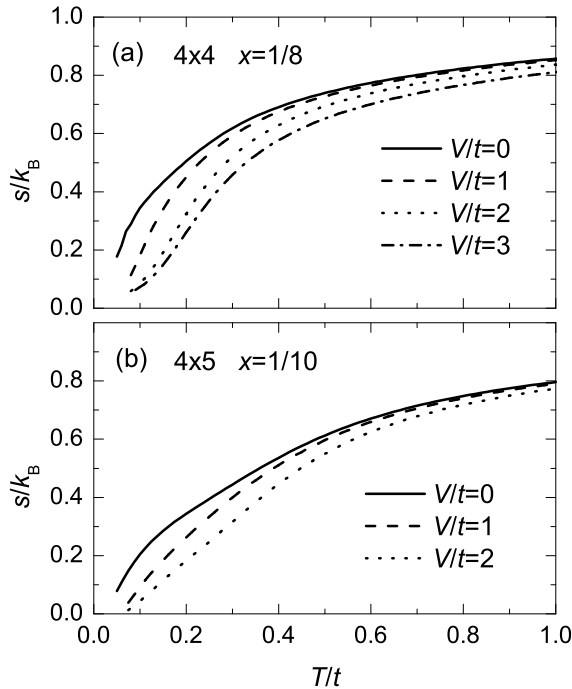}
\caption{\label{fig1}
Entropy density $s$ versus $T$ for different values of the stripe potential $V$.
(a) $4\times 4$ and (b) $4\times 5$ $t$-$J$ clusters with two holes and $J/t=0.4$.}
\end{center}
\end{figure}

First, we examine the dependence of the entropy on the stripe
potential $V$.  The entropy density $s$ for the hole concetration $x$
and temperature $T$ is given by
\begin{equation}
s=\frac{1}{N}\left[ k_\mathrm{B} \ln \Omega + \frac{\langle H_{tJ} +
H_\mathrm{st} \rangle - \mu \left( 1-x \right)N}{T}\right],
\label{s}
\end{equation}
where $k_\mathrm{B}$ is the Boltzmann factor, $\Omega$ is the 
partition function, $\langle \cdots \rangle$ is the averaging within
the ground-canonical ensemble, and $\mu$ is the chemical potential
calculated separately.
Figure~1 shows $s$ as a function of $T$ for $x=1/8$
in the $4\times 4$ cluster and $x=1/10$ in the $4\times 5$
cluster (two holes in both the clusters) at various values of $V/t$ 
in the range of $0\le V/t \le 3$.  Note that at large $V/t$ the two holes are
confined within the stripe,~\cite{Shibata} effectively leading to
quarter-filled 1D charge stripes.  We observe that with increasing $V/t$ $s$ gradually
decreases.  The reduction of $s$ is stronger at low temperatures below
roughly $T\sim J(=0.4t)$ than at high temperatures around $T\sim t$.
Since the decrease of $s$ at high temperatures becomes significant when
$V>t$, the restriction of the space available for the hole motion due to
the stripe potential must be the predominant source of the decrease.
It appears plausible that for $T< J$ AF
spin correlations develop inside the spin domain from which holes are
expelled.~\cite{Shibata} The contribution to the entropy is then small
both from the AF domains without holes (in analogy to the Heisenberg
model where $s \propto T^2$) as well from 1D metallic stripes, unlike
in the homogeneously doped $t$-$J$ model where $s$ remains anomalously
large even at $T<J$.~\cite{Jaklic} The suppression of $s$ in the
stripe phase of La$_{1.6-x}$Nd$_{0.4}$Sr$_x$CuO$_4$ has been indeed
observed experimentally via the small electronic specific heat
coefficient $\gamma$ at $x=1/8$ below $T\sim$30~K.~\cite{Takeda}

\begin{figure}
\begin{center}
\includegraphics[width=8.cm]{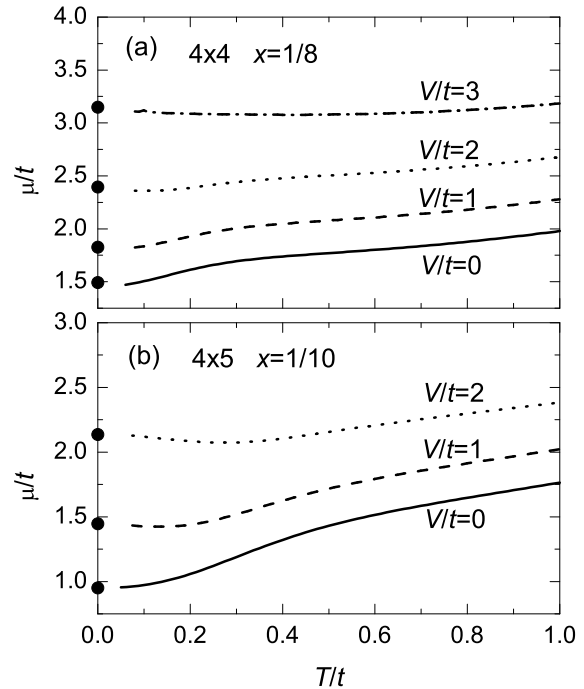}
\caption{\label{fig2}
Chemical potential $\mu$ versus $T$ for different values of the stripe potential $V$.
(a) $4\times 4$ and (b) $4\times 5$ $t$-$J$ clusters with two holes and $J/t=0.4$.
The circles represent $\mu$ at $T=0$.
}
\end{center}
\end{figure}

Next, let us consider the thermoelectric power $Q$ in the stripe phase.  
Since the experimental data mentioned above are just an average from
the contributions parallel and perpendicular to the stripe, it is necessary to take
an average of $Q$ over the two directions to compare with the experiments. 
Within the linear response theory, $Q$ with the electrical and heat
currents in the $\alpha$ direction can be expressed in terms of the current-current
correlation function $C_{j_\alpha j_\alpha}\left(\omega\right)$ and 
the energy current-current correlation function 
$C_{j_{E_\alpha} j_\alpha}\left(\omega\right)$,
\begin{equation}
Q=\frac{1}{eT}\left[\mu-\frac{C_{j_{E_\alpha} j_\alpha}\left(\omega\to
0\right)}{C_{j_\alpha j_\alpha}\left(\omega\to 0\right)}\right],
\label{Qdef}
\end{equation}
where $\mu$ is the chemical potential, and $e$ is the electric charge.  
Instead of a complete analysis of $C_{j_{E_\alpha} j_\alpha}\left(\omega\right)$,
which is straightforward but significantly involved, we rely on an approximate
relation which has been verified within the $t$-$J$ model in the
regime of weak to moderate doping that $C_{j_{E_\alpha} j_\alpha}\left(\omega\right)$
is proportional to $C_{j_\alpha j_\alpha}\left(\omega\right)$.~\cite{Jaklic}
This leads to a simplified expression for the thermoelectric power
\begin{equation}
Q\sim\frac{1}{eT}\left[\mu\left(T\right)-\mu\left(T=0\right)\right].
\label{Qsim}
\end{equation}
Here we note that this approximation removes the difference of $Q$ 
along the parallel and perpendicular directions, since $\mu\left(T\right)$
has no anisotropy.  As a result of this, we do not need to take the average.
As shown below, the experimental features of $Q$ in the
stripe phase is explained by using Eq.~(\ref{Qsim}), indicating the validity 
of the approximation.  More precise calculations of Eq.~(\ref{Qdef}) remains
 to be done in future.
 
\begin{figure}
\begin{center}
\includegraphics[width=8.cm]{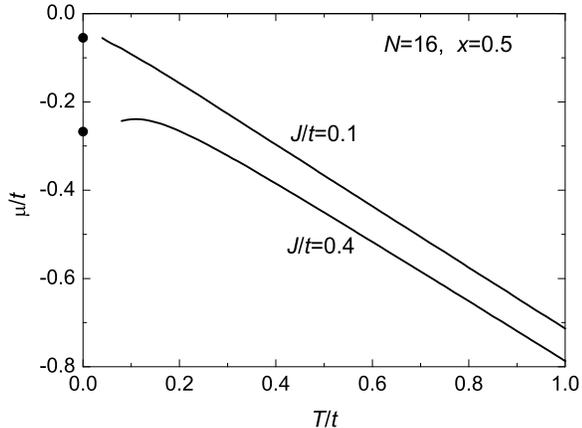}
\caption{\label{fig3}
Chemical potential $\mu$ versus $T$ for a 16-site 1D $t$-$J$ ring with eight holes
($x$=0.5).   $J/t=$0.4 and 0.1.  The circles represent $\mu$ at $T=0$.
}
\end{center}
\end{figure}

We examine $\mu$ before discussing $Q$.
Figure~2 shows $\mu(T)$ for various values of $V/t$ in the two clusters
with two holes.  The chemical potential at $T=0$ is calculated by using
the following equation:
$\mu(T=0)=\left[ E_0\left( 1 \right) - E_0\left( 3 \right) \right]/2$,
where $E_0 \left( N \right)$ is the ground state energy of the $N$-hole
system.  At $V/t=0$, $\mu$ increases with increasing $T$, being
consistent with the previous report that the slope is positive below
$x\sim$0.15.~\cite{Jaklic}  While the slopes at high temperatures near
$T\sim t$ do not change so much when $V/t$ is turned on, those at low
temperatures below $T\sim J$ become smaller and change the sign from positive
to negative at large $V$ [$V/t=$3 and 2 in Figs.~2(a) and 2(b), respectively].
The appearance of a negative slope of $\mu$
again has to be related to the formation of the 1D charge river.  To establish this,
we show in Fig.~3 the temperature dependence of $\mu$ for a 16-site 1D
$t$-$J$ ring with 8 holes (quarter filling at $x=0.5$).  We take an antiperiodic
boundary condition in order to obtain the nondegenerate ground state.
We find that the slope of $\mu$ for $J/t$=0.4 is negative when $T/t\gtrsim$0.15.
The negative slope at around $T/t=$0.2 is consistent with the data at large $V$
in Fig.~2.  In Fig.~2, the negative slope region extends to lower temperatures
when the data are extrapolated to the $T=$0 values.  This seems to be different
from the temperature dependence of $\mu$ for $J/t$=0.4 below $T/t\sim$0.15
in Fig.~3.  However, with reducing $J/t$, we find that the negative slope
region extends to lower temperatures (see $\mu$ for $J/t$=0.1), indicating
qualitatively similar behavior with the data at large stripe potential.
Therefore, it is as likely that the value of $J/t$ is effectively
reduced inside the charge stripes.  Actually, it may be possible that
the coupling of spins in the stripes with those in the AF spin domains
induces frustration effects that effectively reduce the exchange
interaction inside the stripes.  This possibility should be examined
in a detail in the future.

\begin{figure}
\begin{center}
\includegraphics[width=8.cm]{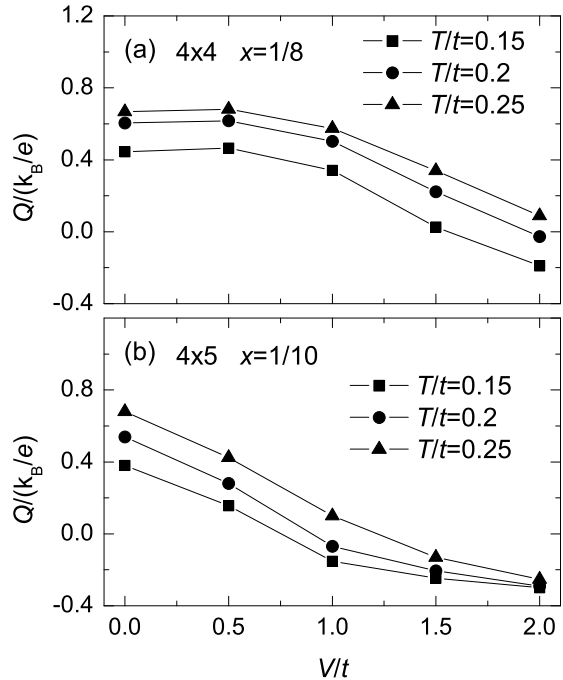}
\caption{\label{fig4}
Thermoelectric power $Q$ versus stripe potential $V$ at several temperatures.
(a) $4\times 4$ and (b) $4\times 5$ $t$-$J$ clusters with two holes and $J/t=0.4$.
$Q$ is evaluated by using the expression Eq.~(\ref{Qsim}).
}
\end{center}
\end{figure}

According to Eq.~(\ref{Qsim}), we evaluate $Q\left(T\right)$, which is shown in
Fig.~4.  At the temperatures shown in the figure, $Q$ decreases with increasing $V$.
This means that the formation of the stripe phase reduces $Q$, being consistent with
experimental data that $Q$ is strongly suppressed in the stripe phase of La-based
cuprates.  Fixing the value of $V$, $Q$ decreases with decreasing temperature,
being again consistent with the experimental data below the LTO-LTT transition
temperature.  At large $V$, $Q$ becomes negative.  As discussed above, this
may be associated with the presence of the quarter-filled 1D system with strong
correlations.  Very interestingly, negative $Q$ is obtained in such a
quarter-filled system when $T$ is much larger than $t$ but much smaller than
the on-site Coulomb repulsion $U$.  In this limit, $Q$ is given by
$\mu/\left( eT\right)$, leading to $Q=-\left( k_\mathrm{B}/e \right)\ln
\left\{ 2x/\left(1-x\right) \right\}=-\left( k_\mathrm{B}/e \right)\ln 2$
at $x$=0.5.~\cite{Chaikin}  Since the negative slope of $\mu$ in the quarter-filled 1D
$t$-$J$ model shown in Fig.~3 is smoothly connected to the value
$-k_\mathrm{B}\ln 2$ at $T\rightarrow \infty$, the negative $Q$ at
large $V$ in Fig.~4 may be the realization of characteristics in the
quarter-filled system with strong correlations.  From Fig.~4 we can
also conclude that in a broader $4\times 5$ system
$Q$ decreases faster with imposed potential $V/t$. This seems
plausible since one expects that in a larger lattice,
in particular with open boundary conditions, the $t$-$J$ model
tends inherently towards stable stripe structures,~\cite{White}
hence generally a weaker $V/t$ is enough to stabilize stripes,
although the values of $V/t$ taken here are still large probably due to 
small system sizes.

Based on the results above, we further discuss the physical meaning of
the reduction of $Q$ in the stripe phase.  The temperature derivative of
$\mu$ is related to the density derivative of $s$ according to
the Maxwell's relation
\begin{equation}
\left(\frac{\partial\mu}{\partial T}\right)_x=\left(\frac{\partial s}{\partial
x}\right)_T.
\label{Max}
\end{equation}
The small temperature dependence of $\mu$ in the presence of the stripe potential
shown in Fig~2 implies small variation of the entropy at around $x=1/8$.  In
fact, $s$, which is an increasing function with respect to $x$ at around
$x=1/8$ for $V=0$,~\cite{Jaklic} is suppressed near $x=1/8$ with increasing $V$,
leading to the small value of $\left(\partial s/\partial
x\right)_T$.  Therefore, the reduction of $Q$ induced by the
small temperature dependence of $\mu$ is intimately related to the small variation
of the entropy with respect to the hole concentration.  This is reasonable because
holes added to the stripe phase enter into the charge domain wall (stripe)
not affecting AF spin domains and therefore induce only a small change of
the entropy.  We can thus conclude
that the suppression of $Q$ below the LTO-LTT transition is a consequence of the
reduction of $s$ accompanied by small concentration dependence.

In summary, we have examined the entropy, chemical potential, and thermoelectric power in
the stripe phase of high-$T_c$ cuprates, by using the finite-temperature Lanczos
technique for the $t$-$J$ model with a stripe potential.  We find that the
thermoelectric power becomes small when the stripe potential is introduced.
This is understood as a consequence of the reduction of the entropy associated
with the formation of the one-dimensional charge rivers characteristic of the
stripe phase.  The suppression of the thermoelectric power as well as the
entropy is consistent with the experimental data in the stripe phase of
La$_{1.6-x}$Nd$_{0.4}$Sr$_x$CuO$_4$.  From both the preceding~\cite{Prelovsek}
and this works, we conclude that the anomalies of the Hall constant,
thermoelectric power, and entropy in the stripe phase have all the origin
in the formation of the quarter-filled one-dimensional charge stripes.

We would like to thank W. Koshibae for enlightening discussions.
This work was partly supported by a Grant-in-Aid for scientific Research from the
Ministry of Education, Culture, Sports, Science and Technology of Japan, CREST,
and by Ministry of Education, Science and Sports of Slovenia.
The numerical calculations were performed in the supercomputing facilities
in ISSP, University of Tokyo, IMR, Tohoku University, and IJS, Ljubljana.

\end{document}